\documentclass[twocolumn,prl,superscriptaddress]{revtex4}
\usepackage{amsmath}
\usepackage{graphicx}
\usepackage{txfonts}

\begin{document}

\title{Positive phase space transformation incompatible with classical
physics}
\author{Wonmin Son}
\affiliation{The School of Physics and Astronomy, University of Leeds, LS2 9JT Leeds,
United Kingdom}
\affiliation{Centre for Quantum Technologies, National University of Singapore, 3 Science
Drive 2, Singapore 117543, Singapore}
\affiliation{Department of Physics, National University of Singapore, 2 Science Drive 3,
Singapore 117543, Singapore}
\author{Johannes Kofler}
\affiliation{Fakult\"at f\"ur Physik, Universit\"at Wien, Boltzmanngasse 5, A--1090 Wien,
Austria}
\affiliation{Institut f\"ur Quantenoptik und Quanteninformation, \"Osterreichische
Akademie der Wissenschaften, Boltzmanngasse 3, A--1090 Wien, Austria}
\author{M. S. Kim}
\affiliation{School of Mathematics and Physics, The Queen's University, Belfast BT7 1NN,
United Kingdom}
\author{Vlatko Vedral}
\affiliation{The School of Physics and Astronomy, University of Leeds, LS2 9JT Leeds,
United Kingdom}
\affiliation{Centre for Quantum Technologies, National University of Singapore, 3 Science
Drive 2, Singapore 117543, Singapore}
\affiliation{Department of Physics, National University of Singapore, 2 Science Drive 3,
Singapore 117543, Singapore}
\author{{\v C}aslav Brukner}
\affiliation{Fakult\"at f\"ur Physik, Universit\"at Wien, Boltzmanngasse 5, A--1090 Wien,
Austria}
\affiliation{Institut f\"ur Quantenoptik und Quanteninformation, \"Osterreichische
Akademie der Wissenschaften, Boltzmanngasse 3, A--1090 Wien, Austria}
\date{\today }

\begin{abstract}
Bell conjectured that a positive Wigner function does not allow violation of
the inequalities imposed by local hidden variable theories. A requirement
for this conjecture is "when phase space measurements are performed". We
introduce the theory-independent concept of "operationally local
transformations" which refers to the change of the switch on a local
measurement apparatus. We show that two separated parties, performing only
phase space measurements on a composite quantum system with a positive
Wigner function and performing only operationally local transformations that
preserve this positivity, can nonetheless violate Bell's inequality. Such
operationally local transformations are realized using entangled ancillae.
\end{abstract}

\maketitle

Quantum theory makes only probabilistic predictions in the description of
nature. Since the seminal gedanken experiment by Einstein, Podolsky and
Rosen (EPR) \cite{Einstein35}, there had been a long debate whether one can
go beyond probabilities and arrive at a deterministic description of nature.
Bell has shown that theories which base on the two assumptions of \textit{%
realism} and \textit{locality} have to fulfill certain inequalities \cite%
{Bell87}. Realism assumes the existence of "hidden variables" which
determine the properties of physical systems prior to and independent of
measurements. Locality assumes that these properties cannot be influenced by
space-like separated events. Bell's inequalities are violated by quantum
mechanical predictions for entangled states, and under reasonable
assumptions numerous experiments have disproved local realism. Entanglement
also plays a decisive role in the new field of quantum information, where
quantum correlations are exploited to perform classically impossible tasks
(e.g., quantum cryptography and quantum communication complexity \cite%
{Niel2000}).

To construct a local hidden variable model for correlation measurements on
two particles, it is necessary to find a positive probability distribution
over their local hidden variables. One might think that such a model is
constructed by finding a representation of the state in terms of a positive
probability distribution of phase space variables (position and momentum).
However, it has been argued that the existence of such a distribution is not
a sufficient condition for satisfying Bell's inequality \cite{Bana98}.
Rather, one has to also restrict oneself to certain measurement observables
(phase space measurements).

We will explain the necessary conditions for the existence of a local hidden
variable model using the Wigner quasi probability distribution (which can be
negative in general) and the historic example of the (non-normalized)
entangled EPR state $\int_{-\infty }^{\infty }$d$x\,\left\vert
x\right\rangle _{1}\left\vert x\!+\!x_{0}\right\rangle _{2}=\int_{-\infty
}^{\infty }$d$p\,$exp$(\tfrac{\text{i}}{\hbar }x_{0}p)\,\left\vert
p\right\rangle _{1}\left\vert -p\right\rangle _{2}$. Here, $x$ and $p$
denote position and momentum, $x_{0}$ is a constant, and indices label the
two particles. Bell conjectured that this state does not violate Bell's
inequality imposed by local hidden variables due to the fact that---although
the state is entangled (non-separable)---its Wigner function \cite{Wigner32}
$W(x_{1},p_{1},x_{2},p_{2})=\delta (x_{1}\!-\!x_{2}\!+\!x_{0})\,\delta
(p_{1}\!+\!p_{2})$ is positive in the whole phase space \cite{Bell87}. Here,
$x_{1},p_{1},x_{2},p_{2}$ can serve as local hidden variables with $W$ being
a positive probability distribution, and correlations can be computed via
suitable integration over the Wigner function. Accordingly, no Bell
inequality can be violated. The EPR states can be experimentally realized
e.g.\ via two-mode squeezed states \cite{Bana99a}.

While Bell's conjecture is true for phase space measurements, it has been
shown that nevertheless there are certain other measurements for which local
realism can be violated although the state has a positive Wigner function
\cite{Bana98,Munro99,Chen02}. (One example is the parity measurement of the
number of photons in the two-mode squeezed state.) The reason is that "[t]he
Wigner representations of quantum observables cannot be in general
interpreted as phase space distributions of possible experimental outcomes.
[...] This enables violation of Bell's inequalities even for quantum states
described by positive-definite Wigner functions." \cite{Bana98}

Consider the expectation value of correlations of outcomes $a$ and $b$ for
given local measurements,%
\begin{equation*}
E=\sum a\,b\,p(a,b),
\end{equation*}%
where $p(a,b)$ is the joint probability for the respective outcomes. This
correlation can in general be obtained by
\begin{equation*}
E=\int \text{d}x_{1}\text{d}p_{1}\text{d}x_{2}\text{d}p_{2}\,A(x_{1},p_{1})%
\,B(x_{2},p_{2})\,W(x_{1},p_{1},x_{2},p_{2})
\end{equation*}%
where $A(x_{1},p_{1})$ and $B(x_{2},p_{2})$ are the Wigner representations
of the measurement observables. From the general framework introduced in~%
\cite{Loub08}, it follows that, in case of positive $W$ and bounded $%
A(x_{1},p_{1})$, $B(x_{2},p_{2})$, the corresponding correlations admit a
local realistic description and, therefore, satisfy any correlation Bell
inequality (possibly rescaled) for any numbers of settings and outcomes at
each site. The situation is, however, quite different whenever the functions
$A(x_{1},p_{1})$, $B(x_{2},p_{2})$ are unbounded. For example, in Ref.~\cite%
{Bana98}, the parity measurement outcomes $a$, $b$ can have only values $+1$
or $-1$, but the Wigner representations $A(x_{1},p_{1})$, $B(x_{2},p_{2})$
are given by delta functions. The latter makes the Bell inequality void~\cite%
{Bana98} even in the presence of positive $W$.

It is still questioning whether the positivity of the Wigner function and
using phase space observables are sufficient ingredients to claim whether
the system is genuinely classical. This is an important question since it
might be related to the recent dispute on the power of nuclear magnetic
resonance quantum computation. It is claimed that a system in a highly noisy
statistical mixture can be used as a resource for quantum computation \cite%
{Schack99}. Although all states involved in the experiment have a local
realistic model, one can exploit the fact that the transformations between
the states are essentially quantum mechanical, not having any classical
description \cite{Lind2001}. This implies that the describability of a
prepared system in terms of local realism may not rule out its usefulness as
a resource for quantum information processing.

Throughout the paper, we will use the term "local transformation" in an
\textit{operational} meaning. It is defined theory-independently and refers
to the \textit{change of a switch on a local measurement apparatus}. A
general measurement at one observer side can be understood as a
concatenation of a corresponding (operationally) local state transformation
and a fixed measurement (in the computational basis). Naively, one would
assume that if one has a local hidden variable model for a quantum state and
one applies only operationally local transformations, no Bell inequalities
can be violated. We will show that this is wrong.

We present an explicit example of a quantum state with a positive Wigner
function and operationally local transformations that keep this positivity,
as well as a fixed phase space measurement on each side, which nevertheless
allows to violate Bell's inequality. This demonstrates that even the
conjunction of (i) positivity of the system's Wigner function, (ii) fixed
phase space measurements, and (iii) operationally local transformations
preserving the positivity of the system's Wigner function is not sufficient
to claim the possibility of a local realistic model. The trick here is that
although the quantum state itself has a local hidden variable model, the
used operationally local transformations are completely positive (CP) maps
\textit{which can only be simulated in the hidden variable model via
non-local transformations}. Such a situation is achieved by exploiting
additional entangled particles (called "ancillae" in quantum information
theory).

We will now derive a Bell inequality in terms of local phase space
transformations in local realistic theories, and our results can therefore
be understood as a "Bell theorem for local transformations". These theories
are based on two postulates: (I)~Realism:\ Systems have well-defined
properties prior to and independent of measurement. (II)~Locality:\ The
properties on one side are independent of any (space-like) separated
events---in particular of operationally local transformations---on the other
side.

Consider two separated observers, usually called Alice and Bob, who measure
the phase space observables $A_{i}$ and $B_{j}$ ($i,j=1,2$) of a system with
positive Wigner function $W$. The expectation value of the correlation reads%
\begin{equation}
\langle A_{i}B_{j}\rangle _{W}=\int \text{d}\mathbf{z}_{1}\text{d}\mathbf{z}%
_{2}\,W(\mathbf{z}_{1},\mathbf{z}_{2})\,A_{i}(\mathbf{z}_{1})\,B_{j}(\mathbf{%
z}_{2}),  \label{eq AiBjW}
\end{equation}%
with $\mathbf{z}_{1}=(x_{1},p_{1})$ and $\mathbf{z}_{2}=(x_{2},p_{2})$ the
shortcuts for the local hidden variables (e.g.\ position and momentum) of
Alice and Bob, respectively. Note that in local phase space measurements the
outcomes depend only on the position and momentum of the local particles.
Eq.~(\ref{eq AiBjW}) represents a local realistic model for phase space
measurements on a bipartite quantum state with a positive Wigner function,
or the classical average of products under phase space measurements on two
classical particles. Taking dichotomic observables $A_{i}$ and $B_{j}$, the
Clauser-Horne-Shimony-Holt (CHSH) inequality \cite{CHSH69} reads%
\begin{equation}
\langle A_{1}B_{1}\rangle _{W}+\langle A_{1}B_{2}\rangle _{W}+\langle
A_{2}B_{1}\rangle _{W}-\langle A_{2}B_{2}\rangle _{W}\leq 2.
\end{equation}%
The correlation $\langle A_{i}B_{j}\rangle _{W}$ can also be established by
performing \textit{operationally local transformations}, which---\textit{%
within local realistic theories}---have the form%
\begin{equation}
\mathbf{z}_{1}\rightarrow \mathbf{z}_{1}^{(i)}=\mathbf{S}_{i}(\mathbf{z}%
_{1}),\quad \mathbf{z}_{2}\rightarrow \mathbf{z}_{2}^{(j)}=\mathbf{T}_{j}(%
\mathbf{z}_{2}),  \label{eq Ti}
\end{equation}%
followed by measuring fixed phase space observables (no index) $A(\mathbf{z}%
_{1}^{(i)})=A_{i}(\mathbf{S}_{i}^{-1}(\mathbf{z}_{1}^{(i)}))$, $B(\mathbf{z}%
_{2}^{(j)})=B_{j}(\mathbf{T}_{j}^{-1}(\mathbf{z}_{2}^{(j)}))$:%
\begin{equation}
\langle AB\rangle _{W_{ij}}=\int \text{d}\mathbf{z}_{1}^{(i)}\text{d}\mathbf{%
z}_{2}^{(j)}\,J_{ij}\,W_{ij}(\mathbf{z}_{1}^{(i)},\mathbf{z}_{2}^{(j)})\,A(%
\mathbf{z}_{1}^{(i)})\,B(\mathbf{z}_{2}^{(j)}).  \label{eq ABWij}
\end{equation}%
In eq.\ (\ref{eq ABWij}) $W_{ij}(\mathbf{z}_{1}^{(i)},\mathbf{z}%
_{2}^{(j)})=W(\mathbf{S}_{i}^{-1}(\mathbf{z}_{1}^{(i)}),\mathbf{T}_{j}^{-1}(%
\mathbf{z}_{2}^{(j)}))$ is the Wigner function of the state after the
invertible local transformations $\mathbf{S}_{i}$ and $\mathbf{T}_{j}$, and $%
J_{ij}$ is the determinant of the Jacobian of this transformation. The
equivalence of eqs.~(\ref{eq AiBjW}) and (\ref{eq ABWij}) is closely related
to the equivalence of active and passive transformations. Here it is crucial
that within local hidden variable models the operationally local
transformations (\ref{eq Ti}) are local transformations of the hidden
variables, i.e.\ $\mathbf{z}_{1}^{(i)}=\mathbf{S}_{i}(\mathbf{z}_{1})$ is
not a function of $\mathbf{z}_{2}$ and $\mathbf{z}_{2}^{(j)}=\mathbf{T}_{j}(%
\mathbf{z}_{2})$ is not a function of $\mathbf{z}_{1}$.

Taking again dichotomic observables $A_{i}$ and $B_{j}$, allows to rewrite
the CHSH inequality in the form%
\begin{equation}
\langle AB\rangle _{W_{11}}+\langle AB\rangle _{W_{12}}+\langle AB\rangle
_{W_{21}}-\langle AB\rangle _{W_{22}}\leq 2.  \label{eq CHSH Wij}
\end{equation}

We will show that four positive Wigner functions allow to violate the CHSH
inequality for fixed phase space quantum measurements, even though they
remain positive throughout the transformations. Note that this is different
from choosing four positive arbitrary Wigner functions $f_{ij}$ for which
the above inequality with the terms $\langle AB\rangle _{f_{ij}}$ can always
be violated. The point is that the functions $W_{ij}$ in (\ref{eq CHSH Wij})
are obtained by operationally local transformations.

Consider the mixture of coherent states%
\begin{equation}
\hat{\rho}_{0}=\tfrac{1}{2}\,\Big(\left\vert \alpha ,\alpha \right\rangle
_{12}\!\left\langle \alpha ,\alpha \right\vert +\left\vert -\alpha ,-\alpha
\right\rangle _{12}\!\left\langle -\alpha ,-\alpha \right\vert \Big),
\label{eq:coherent}
\end{equation}%
where $|\alpha \rangle $ denotes a coherent state of complex amplitude $%
\alpha $ \cite{Glauber63} and 1 and 2 label Alice and Bob's particle
respectively. For large $|\alpha |$, the two states $\left\vert \alpha
\right\rangle $ and $\left\vert -\alpha \right\rangle $ become almost
orthogonal as $|\langle \alpha \left\vert -\alpha \right\rangle
\!|^{2}=e^{-4|\alpha |^{2}}\approx 0$. From now on, we assume that $|\alpha
| $ is sufficiently large.

The Wigner function of the state $\hat{\rho}_{0}$, $W_{\hat{\rho}_{0}}(\beta
_{1},\beta _{2})=\tfrac{1}{2}[W_{\alpha }(\beta _{1})W_{\alpha }(\beta
_{2})+W_{-\alpha }(\beta _{1})W_{-\alpha }(\beta _{2})]$, is positive at
every point in phase space because the Wigner function for a single coherent
state $|\alpha \rangle $, i.e.\ $W_{\alpha }(\beta )=\tfrac{2}{\pi }%
e^{-2|\alpha -\beta |^{2}}$, is positive. Here, the complex numbers $\beta
_{1}$ and $\beta _{2}$ take the role of Alice's and Bob's local hidden
variables (as $\mathbf{z}_{1}$ and $\mathbf{z}_{2}$ in eq.~(\ref{eq AiBjW}%
)). Note that all quasi probability distributions for the separable state (%
\ref{eq:coherent}) are positive, and therefore the results of this work are
representation-independent.

In a Bell experiment, Alice and Bob measure the local quadratures (as a
phase space measurement) and assign either the value $+1$ or $-1$ depending
on the sign of the quadrature. Thus, Alice' observable is $\hat{A}(\theta
)\equiv \mbox{Sign}[\hat{x}_{\theta }]$ where $\hat{x}_{\theta }=\cos \theta
\,\hat{x}+\sin \theta \,\hat{p}$ is the quadrature operator (along the angle
$\theta $) with $\hat{x}$ and $\hat{p}$ the position and momentum operator
\cite{Barnett97}. The explicit form of the measurement operator is $\hat{A}%
(\theta )=(\int_{0}^{\infty }$d$x_{\theta }-\int_{-\infty }^{0}$d$x_{\theta
})|x_{\theta }\rangle \langle x_{\theta }|$ where $|x_{\theta }\rangle $ is
eigenstate of the quadrature operator $\hat{x}_{\theta }$. We assume that
Bob measures his system with the same type of measurement $\hat{B}(\varphi )$%
. The correlation between Alice and Bob takes the form $\langle
A_{i}B_{j}\rangle =\mbox{Tr}[\hat{A}(\theta _{i})\hat{B}(\varphi _{j})\hat{%
\rho}_{0}]$ which can be used for the CHSH inequality.

Now we will show that one can violate the CHSH inequality under
operationally local transformations. The initial state (\ref{eq:coherent})
can be transformed into the state%
\begin{equation}
\hat{\rho}(\theta _{i},\varphi _{j})=\cos ^{2}(\theta _{i}\!-\!\varphi
_{j})\,\hat{\rho}_{0}+\sin ^{2}(\theta _{i}\!-\!\varphi _{j})\,\hat{\rho}_{1}
\label{eq:newstate1}
\end{equation}%
with $\hat{\rho}_{1}\equiv \tfrac{1}{2}(\left\vert \alpha ,\!-\alpha
\right\rangle \!\left\langle \alpha ,\!-\alpha \right\vert +\left\vert
-\alpha ,\!\alpha \right\rangle \!\left\langle -\alpha ,\!\alpha \right\vert
)$, and where $\theta _{i}$ and $\varphi _{j}$ denote local parameters for
the operationally local transformations and are controlled by Alice and Bob,
respectively. It is not immediately obvious how this state can be generated
but we will discuss this later. It is straightforward that the positivity of
the Wigner function of the transformed state $\hat{\rho}(\theta _{i},\varphi
_{j})$ is preserved since $W_{ij}(\beta _{1},\beta _{2})=\tfrac{1}{2}\{\cos
^{2}(\theta _{i}\!-\!\varphi _{j})[W_{\alpha }(\beta _{1})W_{\alpha }(\beta
_{2})+W_{-\alpha }(\beta _{1})W_{-\alpha }(\beta _{2})]+\sin ^{2}(\theta
_{i}\!-\!\varphi _{j})[W_{\alpha }(\beta _{1})W_{-\alpha }(\beta
_{2})+W_{-\alpha }(\beta _{1})W_{\alpha }(\beta _{2})]\}$ which has only
positive contributions. As under the transformations the state remains
separable and also with positive Wigner function, one could naively
characterize this transformation as classical.

Contrary to this intuition, we will show that $W_{11}$, $W_{12}$, $W_{21}$,
and $W_{22}$ allow to violate the CHSH inequality (\ref{eq CHSH Wij}). The
correlation function for the state (\ref{eq:newstate1}) with fixed
("position left/right") phase space measurements $\hat{A}\equiv \hat{A}(0)$
and $\hat{B}\equiv \hat{B}(0)$ is $\langle AB\rangle _{W_{ij}}=\mbox{Tr}[%
\hat{A}\,\hat{B}\,\hat{\rho}(\theta _{i},\varphi _{j})]=\cos [2(\theta
_{i}\!-\!\varphi _{j})]\mbox{Erf}^{2}(\alpha _{r})$ with $\alpha _{r}\equiv %
\mbox{Re}(\alpha )$. It is notable that the correlation function cannot be
factorized into individual parts for Alice and Bob. The CHSH expression $%
C(\theta _{1},\theta _{2},\varphi _{1},\varphi _{2})\equiv \langle AB\rangle
_{11}+\langle AB\rangle _{12}+\langle AB\rangle _{21}-\langle AB\rangle
_{22} $ becomes $C(\theta _{1},\theta _{2},\varphi _{1},\varphi
_{2})=(\Delta _{11}+\Delta _{12}+\Delta _{21}-\Delta _{22})\,\mbox{Erf}%
^{2}(\alpha _{r})$ where $\Delta _{ij}\equiv \cos [2(\theta _{i}\!-\!\varphi
_{j})]$. With a proper choice of the local parameters, e.g.\ $(\theta
_{1},\theta _{2},\varphi _{1},\varphi _{2})=(0,\pi /4,\pi /8,-\pi /8)$, the
CHSH expression $C(0,\pi /4,\pi /8,-\pi /8)=2\sqrt{2}~\mbox{Erf}^{2}(\alpha
_{r})$ violates the CHSH inequality when $\alpha _{r}>0.9957$. Thus, we can
conclude that operationally local transformations applied on a quantum state
with a positive Wigner function can violate a Bell inequality, even though
they keep the positivity of the Wigner function. To simulate the violation,
at least one of the transformations between the four Wigner functions
modeled by the hidden variables must be non-local, i.e.\ cannot be written
as in eqs.\ (\ref{eq Ti}) but only as $\mathbf{z}_{1}^{(i)}=\mathbf{S}_{i}(%
\mathbf{z}_{1},\mathbf{z}_{2})$, $\mathbf{z}_{2}^{(j)}=\mathbf{T}_{j}(%
\mathbf{z}_{1},\mathbf{z}_{2})$.

We demonstrate that it is not possible to realize the set of transformations
leading to a violation by local \textit{quantum} operations (allowing
classically correlated ancillae at both sides). Let us first consider the
case where only Alice performs a general operation on her particle and Bob
does nothing on his particle. Then, the transformation from (\ref%
{eq:coherent}) to (\ref{eq:newstate1}) can be described by a positive
operator-sum representation at Alice's side $\mathcal{\hat{L}}_{i}^{(A)}(%
\hat{\rho}_{0})=\sum_{\mu =1}^{2}{\mathcal{\hat{E}}_{\mu }^{i}}\hat{\rho}_{0}%
{\mathcal{\hat{E}}_{\mu }^{i\dag }}$ where $\hat{\rho}_{0}$ is the initial
state in (\ref{eq:coherent}). Since the transformation changes only the
diagonal elements, the most general positive operators for the local
transformation by Alice are ${\mathcal{\hat{E}}}_{1}^{i}=\cos \theta
_{i}(\left\vert \alpha \right\rangle \!\left\langle \alpha \right\vert
+\left\vert -\alpha \right\rangle \!\left\langle -\alpha \right\vert )$ and $%
{\mathcal{\hat{E}}}_{2}^{i}=\sin \theta _{i}(\left\vert \alpha \right\rangle
\!\left\langle -\alpha \right\vert +\left\vert -\alpha \right\rangle
\!\left\langle \alpha \right\vert )$ which describe the bit flip channel of
the qubit and satisfy $\sum_{\mu =1}^{2}{\mathcal{\hat{E}}_{\mu }^{i\dag }%
\mathcal{\hat{E}}_{\mu }^{i}}=\openone$. Neither the phase flip nor the
bit-phase flip operation can produce the transformation as they would
introduce non-diagonal elements~\cite{Niel2000}. After the operation, the
state becomes $\mathcal{L}_{i}^{(A)}(\hat{\rho}_{0})=\hat{\rho}(\theta
_{i},0)$. If Bob performs the same local operation on his particle with $%
\varphi _{j}$ instead of $\theta _{i}$, then the state is transformed into $%
\mathcal{\hat{L}}_{j}^{(B)}(\hat{\rho}_{0})=\rho (0,\varphi _{j})$. However,
the composition of the local operations by Alice and Bob does not transform
the initial state into (\ref{eq:newstate1}), i.e.%
\begin{equation*}
\mathcal{\hat{L}}_{i}^{(A)}\mathcal{\hat{L}}_{j}^{(B)}(\hat{\rho}%
_{0})=\sum_{\mu ,\nu =1}^{2}({\mathcal{\hat{E}}}_{\mu }^{i}\otimes {\mathcal{%
\hat{E}}}_{\nu }^{j})\,\hat{\rho}_{0}\,({\mathcal{\hat{E}}_{\mu }^{i\dag }}%
\otimes {\mathcal{\hat{E}}_{\nu }^{j\dag }})\neq \hat{\rho}(\theta
_{i},\varphi _{j})
\end{equation*}%
for arbitrary local parameters $\theta _{i}$ and $\varphi _{j}$. This shows
that the initial state $\hat{\rho}_{0}$, eq.\ (\ref{eq:coherent}), cannot be
transformed into the state $\hat{\rho}(\theta _{i},\varphi _{j})$, eq.\ (\ref%
{eq:newstate1}), by any local quantum operations possibly assisted by
classically correlated ancillae.

We now show how the state transformation can be obtained by local quantum
operations assisted by entangled ancillae. Note that such a quantum
operation is an operationally local transformation. Let us assume that in
addition to the initial state $\hat{\rho}_{0}$ Alice and Bob share an
entangled state $|\psi \rangle _{34}=\frac{1}{\sqrt{2}}(\left\vert
00\right\rangle _{34}+\left\vert 11\right\rangle _{34})$ where the
subscripts $3$ and $4$ denote the qubit ancillae belonging to Alice and Bob,
respectively. Then, Alice rotates her original particle (label 1) together
with her ancilla (label 3) by the following local unitary operation%
\begin{equation}
\hat{U}_{13}(\theta _{i})=%
\begin{pmatrix}
\cos \theta _{i} & -\sin \theta _{i} & 0 & 0 \\
0 & 0 & \sin \theta _{i} & \cos \theta _{i} \\
0 & 0 & \cos \theta _{i} & -\sin \theta _{i} \\
\sin \theta _{i} & \cos \theta _{i} & 0 & 0%
\end{pmatrix}%
\end{equation}%
where the matrix basis is $\{\left\vert \alpha ,0\right\rangle _{13}$, $%
\left\vert \alpha ,1\right\rangle _{13}$, $\left\vert -\alpha
,0\right\rangle _{13}$, $\left\vert -\alpha ,1\right\rangle _{13}\}$. Bob
applies the similar unitary $\hat{U}_{24}(\varphi _{j})$ to his particle
(label 2) and ancilla (label 4). Then the initial state $\hat{\rho}_{0}$ is
transformed to%
\begin{equation}
\mbox{Tr}_{34}[\hat{U}_{13}(\theta _{i})\hat{U}_{24}(\varphi _{j})\,\hat{\rho%
}_{0}\,|\psi \rangle _{34}\langle \psi |\,\hat{U}_{13}^{\dagger }(\theta
_{i})\hat{U}_{24}^{\dagger }(\varphi _{j})]=\hat{\rho}(\theta _{i},\varphi
_{j})  \label{eq olt}
\end{equation}%
where $\mbox{Tr}_{34}$ denotes the partial trace over the ancillae. (In
general, operationally local transformations are represented in quantum
mechanics as in eq.\ (\ref{eq olt}) using CP-maps instead of unitaries as
well as general initial states of the system and the ancillae.) Thus the
required transformation and the violation of the CHSH inequality can be
achieved via operationally local transformations only if they involve
entangled ancillae. Note that the measurements are performed on the
particles 1 and 2 only and not the ancillae.

\begin{figure}[t]
\begin{center}
\includegraphics{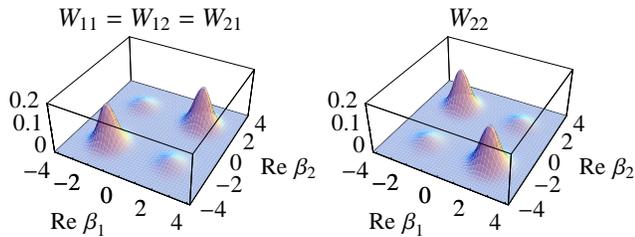}
\end{center}
\par
\vspace{-0.5cm}
\caption{(Color online.) Wigner functions $W_{ij}$ of the density matrix $%
\hat{\protect\rho}(\protect\theta _{i},\protect\varphi _{j})$, eq.\ (\protect
\ref{eq:newstate1}), for the transformations ($\protect\theta _{i},\protect%
\varphi _{j}$) with $i,j=1,2$ and $(\protect\theta _{1},\protect\theta _{2},%
\protect\varphi _{1},\protect\varphi _{2})=(0,\protect\pi /4,\protect\pi /8,-%
\protect\pi /8)$. The $W_{ij}$ are drawn for $\protect\alpha =2$ as a
function of the real parts of Alice's and Bob's hidden variables $\protect%
\beta _{1}$ and $\protect\beta _{2}$, respectively.}
\label{fig:Wigner}
\end{figure}

In Fig.\ \ref{fig:Wigner}, the Wigner functions $W_{ij}(\beta _{1},\beta
_{2})$ for the state (\ref{eq:newstate1}) are plotted (as a function of the
real parts of $\beta _{1},\beta _{2}$), for the transformations ($\theta
_{i},\varphi _{j}$) with $i,j=1,2$ and $(\theta _{1},\theta _{2},\varphi
_{1},\varphi _{2})=(0,\pi /4,\pi /8,-\pi /8)$. We chose $|\alpha |=2$ which
corresponds to an overlap $|\langle \alpha \left\vert -\alpha \right\rangle
\!|^{2}\approx 10^{-7}$. All four Wigner functions are positive and thus
individually allow a local realistic description of phase space
measurements. Obvious local transformations of the hidden variables ($\beta
_{1},\beta _{2}$) that bring $W_{11}$ to $W_{12}$ and $W_{11}$ to $W_{21}$
are identities. Since then, local realistically, the transformation from $%
W_{11}$ to $W_{22}$ has to be the identity as well, all four Wigner
functions would have to be the same. However, quantum mechanically, $W_{22}$
is different from the other functions. The violation of the CHSH inequality (%
\ref{eq CHSH Wij}) for the position left/right measurement shows that no
\textit{local} transformations of the hidden variables exist that can
reproduce all four Wigner functions.

It is usually assumed that no violation of Bell's inequality is possible for
composite systems with positive Wigner function and phase space
measurements. This is because the Wigner function itself can serve as a
probability distribution of local hidden variables. Here we showed that this
statement should be taken with care. Implementing transformations that keep
the positivity of the Wigner function of a composite system at all times, we
have demonstrated that a Bell inequality can be violated performing phase
space measurements. This is possible by exploiting entangled ancillae for
the transformations and the total Wigner function of the composite system
and the ancillae is negative. The important point is that the
transformations---even if they are assisted with entangled ancillae---are
\textit{operationally local} in the sense that they are implemented in a
laboratory by a switch on a local measurement apparatus. If one wants to
model such operationally local transformations in a hidden variable model,
one would require non-local transformations of the hidden variables.

We thank J.~Dunningham for discussions and acknowledge finanical support by
the Erwin Schr\"{o}dinger Institute, Austrian Academy of Sciences, EPSRC,
QIP IRC, Royal Society, Wolfson Foundation, National Research Foundation
(Singapore), Ministry of Education (Singapore), Austrian Science Foundation
FWF (Project No.\ P19570-N16 and SFB), European Commission through Project
QAP (No.\ 015846), and FWF Doctoral Program CoQuS.

\end{document}